\documentclass[twocolumn,  
showpacs,  
preprintnumbers,  
aps,  
prd,  
a4paper,  
nofootinbib,  
tightenlines,  
floats, floatfix  
]{revtex4}

\usepackage{amsmath}
\usepackage{amsfonts}
\usepackage{graphicx,color}

\newcommand{\nn}{\nonumber \\}
\newcommand{\bea}{\begin{eqnarray}}
\newcommand{\ena}{\end{eqnarray}}

\newcommand{\hs}[1]{\hspace{#1 mm}}

\newcommand{\OO}{{\cal O}}
\renewcommand{\a}{\alpha}

\renewcommand{\d}{\delta}

\renewcommand{\O}{\Omega}

\renewcommand{\l}{\lambda}

\begin{document}


\title{Cosmological constraints on Lorentz invariance violation in the neutrino sector}

\author{Zong-Kuan Guo}
\email{guozk@itp.ac.cn}
\affiliation{State Key Laboratory of Theoretical Physics, Institute of Theoretical Physics,
Chinese Academy of Sciences, P.O. Box 2735, Beijing 100190, China}

\author{Qing-Guo Huang}
\email{huangqg@itp.ac.cn}
\affiliation{State Key Laboratory of Theoretical Physics, Institute of Theoretical Physics,
Chinese Academy of Sciences, P.O. Box 2735, Beijing 100190, China}

\author{Rong-Gen Cai}
\email{cairg@itp.ac.cn}
\affiliation{State Key Laboratory of Theoretical Physics, Institute of Theoretical Physics,
Chinese Academy of Sciences, P.O. Box 2735, Beijing 100190, China}

\author{Yuan-Zhong Zhang}
\email{zyz@itp.ac.cn}
\affiliation{State Key Laboratory of Theoretical Physics, Institute of Theoretical Physics,
Chinese Academy of Sciences, P.O. Box 2735, Beijing 100190, China}

\date{\today}

\begin{abstract}
We derive the Boltzmann equation in the synchronous gauge for massive
neutrinos with a deformed dispersion relation.
Combining the 7-year WMAP data with lower-redshift measurements of
the expansion rate, we give constraints on the deformation parameter
and find that the deformation parameter is strong degenerate with
the physical dark matter density rather than the neutrino mass.
Our results show that there is no evidence for Lorentz invariant
violation in the neutrino sector.
The ongoing Planck experiment could provide improved constraints
on the deformation parameter.
\end{abstract}

\pacs{14.60.St, 98.80.Es }

\maketitle

\section{Introduction}

Neutrino oscillations imply that at least two of the three neutrino
types have non-zero mass (see Ref.~\cite{gon07} and
reference therein). Unfortunately, neutrino oscillation experiments
only provide two mass squared differences, but not the overall mass
scale. Cosmology provides a promising way to tackle this problem by
the gravitational effect of massive neutrinos on the expansion
history near the epoch of matter-radiation equality and on the
growth of large-scale structures at late time. Precision
measurements of the cosmic microwave background (CMB) anisotropies
and the large-scale structure distribution of matter have allowed us
to determine or constrain the absolute neutrino mass
scale~\cite{kom08,kom10,hu98,rin04} (see also Ref.~\cite{won11}
for a recent review).

Indeed, neutrino oscillations can be explained by small
Lorentz invariance violation even without introducing
neutrino mass, as shown in~\cite{col97,col98,kos03,kos04,mot12}.
Observed neutrino oscillations may originate from a combination
of effects involving neutrino masses and Lorentz invariance
violation.
The possibilities of Lorentz invariance violation were
explored in quantum gravity~\cite{ame02}, loop quantum
gravity~\cite{alf99}, non-commutative field theory~\cite{car01},
and doubly special relativity theory~\cite{mag01}.

It is well-known that Lorentz symmetry is a fundamental feature of
modern descriptions of the nature, including both the Einstein's
general relativity and standard model of particle physics. One can
expect that the breaking of Lorentz symmetry may leave some imprints
in astrophysical observations such as the CMB anisotropies and
large-scale structure of our universe. In this paper, we consider
cosmological tests of Lorentz invariance violence in the neutrino
sector. Here we focus on the Coleman-Glashow model, in which the
energy-momentum relation is modified by a Lorentz-violating
interaction in the framework of conventional quantum field
theory~\cite{col98}. We will build up a Lagrangian and derive the
Boltzman equation in the synchronous gauge for the massive neutrinos
with deformed dispersion relation, and then place constraints on
Lorentz invariance violation by combining the 7-year WMAP
data~\cite{kom10} with the latest distance measurements from the
Baryon Acoustic Oscillation (BAO) in the distribution of
galaxies~\cite{per09} and the Hubble constant ($H_0$)
measurement~\cite{rie09}.

This paper is organized as follows. In Sec.~\ref{sec2} we write down the
Lagrangian for neutrinos with deformed dispersion relation.
In Sec.~\ref{sec3} we derive the Boltzmann equation for neutrinos in the synchronous
gauge. In Sec.~\ref{sec4} we place constraints on the deformation
parameters using the CMB data in combination with measurements of the
Hubble constant $H_0$ and the BAO feature.
Section~\ref{sec5} is devoted to conclusions.

\section{Deformed dispersion relation}
\label{sec2}

At a phenomenological level the deformed dispersion relation
for massive neutrinos can be generally parameterized by
\bea
E^2 = m^2 + p^2 + \sum_{n=1}^{\infty}\alpha_n \frac{p^n}{M^{n-2}} ,
\label{ddr0}
\ena
where $E$ is the neutrino energy, $m$ is the neutrino mass,
$p=(p^ip_i)^{1/2}$ the magnitude of the 3-momentum,
$\alpha_n$'s are dimensionless coefficients and $M$ denotes the energy scale corresponding to Lorentz symmetry violation  (which is typically taken to be the Planck mass).
Such a deformed dispersion relation implies that there are
departures from Lorentz invariance in the neutrino sector
if $\a_n \neq 0$.
The $n=1$ term would produce huge effects at low energy
and has been strongly constrained.
The $p^n$ term with $n \ge 3$ is suppressed by $1/M^{n-2}$.
In the present work we will therefore consider the case of $n=2$,
\bea
E^2 = m^2 + p^2 + \xi \, p^2,
\label{ddr}
\ena
where $\xi$ is the deformation parameter characterizing the
size of Lorentz symmetry violation.
Such a deformed dispersion relation was constructed by Coleman
and Glashow in the framework of conventional quantum field
theory~\cite{col98}.

Here we point out that the dispersion relation given in
(\ref{ddr0}) and (\ref{ddr}) is not very general.
It neglects oscillations, possible species dependence,
anisotropies associated with violations of rotation symmetry,
and CPT violation.
As shown recently by Kostelecky and Mewes, all of these
are possible~\cite{kos11}.
For example, it is known that taking odd values of $n$ in
Eq.~(\ref{ddr0}) corresponds to CPT violation and gives rise
to a sign difference for neutrinos and antineutrinos~\cite{kos11}.
We have to emphasize that the model considered in this
paper is one of many possible Lorentz-violating theories.

The scalar perturbations of the Friedmann-Lemaitre-Robertson-Walker
metric in the synchronous gauge can be written as
\bea
ds^2=a^2(\tau)\left[-d\tau^2+(\delta_{ij}+h_{ij})dx^i dx^j\right],
\ena
where $a(\tau)$ is the scale factor, $\tau$ is the
conformal time, and the scalar mode of $h_{ij}$ is represented by
two functions $h$ and $\eta$ which are defined by
\bea
h_{ij}(\vec{x},\tau) &=& \int d^3 k e^{i\vec{k} \cdot \vec{x}}
\bigg [\hat{k}_i\hat{k}_jh(\vec{k},\tau)+(\hat{k}_i\hat{k}_j \nn
&& -\frac13\delta_{ij})\;6\eta(\vec{k},\tau)\bigg],
\ena
in Fourier space, where $\vec{k}=k\hat{k}$.
The action for a neutrino with the dispersion relation (\ref{ddr}) can be written in the
first order formalism by
\bea
S=\int d\tau \left[\dot{x}^\mu P_\mu - \l \left(m^2 + g^{\mu\nu}P_\mu P_\nu
 + \xi g^{ij}P_i P_j \right) \right],
\label{fof}
\ena
where $\l$ is a Lagrange multiplier which enforces the mass-shell
condition (\ref{ddr}) and
\bea
P_0 &=& -aE, \\
P_i &=& a(\d_{ij}+\frac12 h_{ij})\,p^j.
\ena
The above action implies that the Hamiltonian vanishes. Performing
the Legendre transformation to the Lagrangian formalism and eliminating
$\l$ from the action (\ref{fof}), we obtain the Lagrangian for
neutrino as follows
\bea
L = -ma\sqrt{1-\frac{u^2}{1+\xi}} \left[1 -
\frac{h_{ij}u^i u^j}{2(1+\xi-u^2)}\right],
\label{lagr}
\ena where
$u^i=dx^i/d\tau$ is the coordinate velocity and $u^2=\d_{ij}u^iu^j$.
Following~\cite{ma93} from the Lagrangian (\ref{lagr}), the
equations of motion in the synchronous gauge become
\bea
\frac{dx^i}{d\tau} &=& \frac{(1+\xi)\d^{ij}q_j}{\epsilon}\left[1+\OO(h)\right]\;,\\
\frac{dq}{d\tau} &=& -\frac12 q \dot{h}_{ij}n^i n^j,
\ena
where $q_j=qn_j$ is the comoving 3-momentum written in terms
of its magnitude and direction with $n^in_i=\d_{ij}n^in^j=1$,
and $\epsilon=\sqrt{m^2a^2+(1+\xi)q^2}$ is the comoving energy of neutrinos.

\section{Boltzmann equation}
\label{sec3}

The number density $n_\nu$, energy density $\rho_\nu$ and
pressure $P_\nu$ for massive neutrinos with (\ref{ddr})
are given by
\bea
\label{nd}
n_\nu &=& \frac{1}{a^4} \int \frac{d^3q}{(2\pi)^3} f_0(q) \;, \\
\rho_\nu &=&
 \frac{1}{a^4} \int \frac{d^3q}{(2\pi)^3} \epsilon f_0(q) \;, \\
P_\nu &=&
 \frac{1}{3a^4} \int \frac{d^3q}{(2\pi)^3} \frac{(1+\xi)q^2}{\epsilon} f_0(q) \;.
\label{pd}
\ena
Here the zeroth-order phase space distribution is
well approximated by the relativistic Fermi-Dirac distribution
$f_0(q)=g_s[1+\exp(\sqrt{1+\xi}\,q/T_0)]^{-1}$, where $T_0$ is the
neutrino temperature today and $g_s=2$ the number of spin degrees of
freedom. If $m \gg T_0$, the total mass of neutrinos $\sum
m=94(1+\xi)^{3/2} \O_\nu h^2$ eV.

The perturbed energy density, pressure, energy flux,
and shear stress in the Fourier space $k$ are respectively given by
\bea
\d \rho_\nu &=&
 \frac{1}{a^4} \int \frac{d^3q}{(2\pi)^3} \epsilon f_0(q) \Psi_0 , \\
\d P_\nu &=&
 \frac{1}{3a^4} \int \frac{d^3q}{(2\pi)^3} \frac{(1+\xi)q^2}{\epsilon} f_0(q) \Psi_0 , \\
(\rho_\nu + P_\nu) \, \theta_\nu &=&
 \frac{k}{a^4} \int \frac{d^3q}{(2\pi)^3} \sqrt{1+\xi}\,q f_0(q) \Psi_1 , \\
(\rho_\nu + P_\nu) \, \sigma_\nu &=&
 \frac{2}{3a^4} \int \frac{d^3q}{(2\pi)^3} \frac{(1+\xi)q^2}{\epsilon} f_0(q) \Psi_2 ,
\ena
where the perturbations $\Psi_l$ satisfy the following
Boltzmann equation
\bea
&& \dot{\Psi}_l + (1+\xi)\frac{q}{\epsilon}
\frac{k}{2l+1}\left[(l+1)\Psi_{l+1}-l\Psi_{l-1}\right] \nn
&& \hs{5}
+ \left(\d_{2l}\frac{1}{15}\dot h + \d_{2l}\frac{2}{5}\dot \eta -
\d_{0l} \frac16 \dot{h}\right)\frac{d\ln f_0}{d \ln q} = 0 \;,
\label{be}
\ena
in the synchronous gauge. Such a Boltzmann hierarchy
is effectively truncated by adopting the following scheme~\cite{ma95}
\bea
\Psi_{l_{\rm max}+1} \approx \frac{(2l_{\rm
max}+1)\epsilon}{(1+\xi)qk\tau}
 \Psi_{l_{\rm max}}-\Psi_{l_{\rm max}-1}.
\ena

The initial conditions for the perturbation $\Psi_l$ and $\eta$
on super-horizon scales ($k\tau \ll 1$) are given by
\bea
\Psi_0 &=& -\frac14 \d_\nu \frac{d\ln f_0}{d\ln q}, \\
\Psi_1 &=& -\frac{\epsilon}{3\sqrt{1+\xi}\,qk} \theta_\nu \frac{d\ln f_0}{d\ln q}, \\
\Psi_2 &=& -\frac{1}{2(1+\xi)} (\sigma_\nu+\frac14 \xi\delta_\nu) \frac{d\ln f_0}{d\ln q}, \\
\eta &=& 2C-\frac{5+9\sqrt{1+\xi}R_\nu-5R_\nu}{6(15+4\sqrt{1+\xi}R_\nu)}C(k\tau)^2,
\ena
where
\bea
\d_\nu &=& -\frac23 C (k\tau)^2, \\
\sigma_\nu &=& \frac{2(2+R_\nu-\sqrt{1+\xi}R_\nu)}{3(15+4\sqrt{1+\xi}R_\nu)}C(k\tau)^2, \\
\theta_\nu &=& -\frac{(23+4R_\nu)\sqrt{1+\xi}}{18(15+4\sqrt{1+\xi}R_\nu)}C(k^4\tau^3).
\ena
Here, $C$ is a dimensionless constant determined by the amplitude
of the fluctuations from inflation and $R_\nu=\rho_\nu/(\rho_\nu+\rho_\gamma)$ during
radiation domination.

\begin{figure}
\begin{center}
\includegraphics[width=80mm]{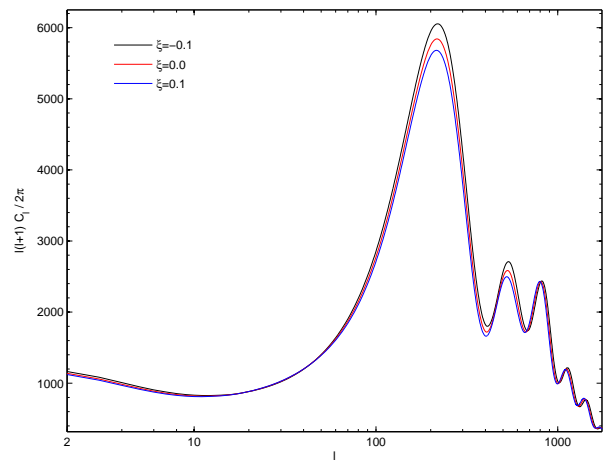}
\caption{Theoretical angular power spectrum of the CMB for
$\xi=-0.1$, $0$, $0.1$.
Here $\sum m=3\times0.3$ eV is fixed.}
\label{fig1}
\end{center}
\end{figure}

In order to compute the theoretical CMB power spectrum, we modified the
Boltzmann CAMB code in~\cite{lew99} to appropriately
incorporate the Lorentz-violating term in the neutrino sector. Actually
the Lorentz-violating term affects not only the evolution
of the cosmological background but also the behavior of the
neutrino perturbations.
From Eqs.~(\ref{nd})-(\ref{pd}), we see that increasing $\xi$
decreases the number density, energy density and pressure of
neutrinos, and thereby increases the redshift when the matter
density equals the radiation density and reduces the expansion
rate prior to and during the epoch of photon-baryon decoupling.
This leads to reduced heights of the first and second peaks of
the CMB while a nearly constant increase in the acoustic
oscillation amplitudes at $l>600$.
On the other hand, the coefficient of the second term in the
Boltzmann equation~(\ref{be}) plays an active role, which
alters the shape of the CMB power spectrum caused by changing
neutrino propagation.
Decreasing $\xi$ increases fluctuation power both at $l<10$ and at $l>100$.
Moreover, the CMB is more sensitive to negative values of $\xi$
than positive ones.
These two effects can be distinguished from a change in the total
mass of neutrinos or in the effective number of extra relativistic
species~\cite{hu95,bas03,ham07,ham10,hou11,guo12},
as shown in Figure~\ref{fig1}.

\section{Cosmological constraints}
\label{sec4}

In our analysis we use a modified version of the publicly
available CosmoMC package to explore the parameter space by
means of Monte Carlo Markov chains technique~\cite{lew02}.
We consider a flat $\Lambda$CDM models plus three Lorentz-violating
neutrino species, described by a set of cosmological parameters
\bea
\{\O_bh^2,\O_ch^2,\Theta_s,\tau,n_s,A_s,\Sigma m,\xi\}, \nonumber
\ena
where $h$ is the dimensionless Hubble parameter such that
$H_0=100h$ kms$^{-1}$ Mpc$^{-1}$, $\O_bh^2$ and $\O_ch^2$ are
the physical baryon and dark matter densities relative to the
critical density,
$\Theta_s$ is  the ratio of the sound horizon to the angular
diameter distance at the photon decoupling, $\tau$ is the reionization
optical depth, $n_s$ and $A_s$ are the spectral index and amplitude
of the primordial curvature perturbations at the pivot scale
$k_0=0.002$ Mpc$^{-1}$, $\sum m$ is the total mass of neutrinos
assuming that the three neutrino masses are approximately degenerate,
and $\xi$ is the deformation parameter.

We use the seven-year WMAP (WMAP7) data with the likelihood
code supplied by the WMAP team.
We consider the Sunyaev-Zel'dovich (SZ) effect, in which CMB
photons scatter off hot electrons in galaxy clusters.
Given a SZ template, the effect is described by a SZ template
amplitude $A_{\rm SZ}$ as in the WMAP paper~\cite{kom10}.
We also impose Gaussian priors on the Hubble constant,
$H_0=74.2 \pm 3.6$ kms$^{-1}$ Mpc$^{-1}$, measured
from the magnitude-redshift relation of low-$z$ type Ia
supernovae~\cite{rie09}, and on the distance ratios,
$r_s/D_V(z=0.2)=0.1905 \pm 0.0061$ and $r_s/D_V(z=0.35)=0.1097 \pm 0.0036$,
measured from BAO in the distribution of galaxies~\cite{per09}.
Here $r_s$ is the comoving sound horizon size at the baryon
drag epoch and $D_V$ is the effective distance measure for
angular diameter distance.

\begin{table*}
\begin{center}
\begin{tabular}{llll}
\hline\hline
parameter & WMAP+$H_0$+BAO & WMAP+$H_0$+BAO & Planck+$H_0$+BAO \\
\hline
$100\Omega_bh^2$ & $2.312\pm0.078$ & $2.314\pm0.081$ & $2.260\pm0.023$  \\
$\Omega_ch^2$    & $0.1200\pm0.0103$ & $0.1206\pm0.0104$ & $0.1133\pm0.0033$  \\
$\Omega_\Lambda$ & $0.715\pm0.020$ & $0.723\pm0.017$ & $0.725\pm0.005$  \\
$H_0$ & $70.9\pm2.4$ km/s/Mpc & $72.1\pm2.2$ km/s/Mpc & $70.3\pm1.0$ km/s/Mpc  \\
$\tau$           & $0.089\pm0.014$ & $0.088\pm0.014$ & $0.088\pm0.004$  \\
$\sum m$         & $<0.51\;{\rm eV}$ & 0 (fixed) & 0 (fixed)  \\
$\xi$            & $-0.077\pm0.089$ & $-0.073\pm0.089$ & $-0.005\pm0.037$ \\
\hline\hline
\end{tabular}
\end{center}
\caption{Mean values and marginalized 68\% confidence level
for the deformation parameter and other cosmological parameters.
For the total mass of neutrinos, the 95\% upper limit is given.}
\label{tab1}
\end{table*}

\begin{figure*}
\begin{center}
\includegraphics[width=160mm]{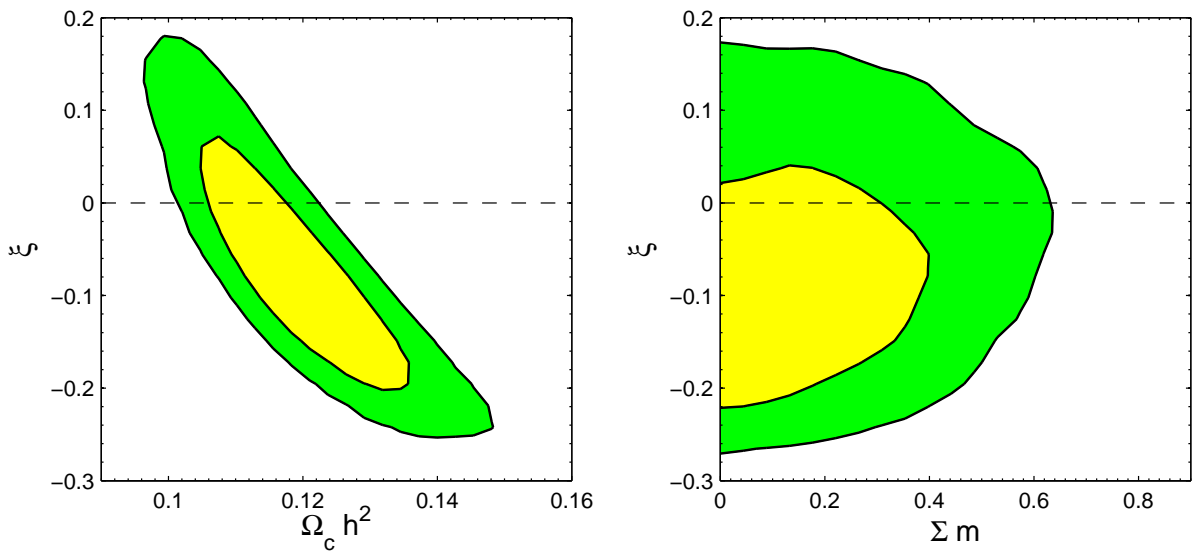}
\caption{Two-dimensional joint marginalized constraints (68\%
and 95\% confidence level) on the deformation parameter $\xi$,
physical dark meter density $\O_ch^2$ (left) and the total mass of
neutrinos $\sum m$ (right), derived from the WMAP7+$H_0$+BAO data.
The dashed line corresponds to Lorentz invariance.}
\label{fig2}
\end{center}
\end{figure*}

In Table~\ref{tab1} we summarize our results.
With the data of WMAP7+$H_0$+BAO, the deformation parameter is
estimated to be $\xi=-0.077\pm0.089$, which implies a null detection
of Lorentz invariant violence within the error limits.
Such large uncertainties in $\xi$ mainly come from the strong
correlation between the deformation parameter and the physical dark
matter density as shown in Figure~\ref{fig2}.
In this case the uncertainties in $\O_c h^2$ are about three times
larger than those derived in the standard $\Lambda$CDM
model~\cite{kom10}.
Since the deformation parameter is nearly uncorrelated with the
total mass of neutrinos in Figure~\ref{fig2}, our constraints on
the deformation parameter are not significantly changed if
three neutrinos are massless, as we can see from Table~\ref{tab1}.
Compared to particle physics experiments, cosmological observations
yield much weaker constraints on the Lorentz-violation parameter
in the neutrino sector.
As listed in Table XIV of Ref.~\cite{kos08}, previous constraints
range from parts in $10^5$ to parts in $10^{15}$ from time-of-flight
measurements and various threshold analyses.

Moreover, we present the constraints on the deformation parameters
from the ongoing Planck experiment~\cite{pla06} in Table~\ref{tab1}.
Following the MCMC method described in Ref.~\cite{per06}, we
generate synthetic data for the Planck experiment and then
perform a systematic analysis on the simulated data.
As we can see in Table~\ref{tab1}, the Planck data plus
measurements of the Hubble constant and the angular diameter distance
will reduce the uncertainties in $\xi$ by a factor of 2.4.
Therefore, Planck CMB measurements allow us to detect the signature
of Lorentz invariant violence at 2$\sigma$ confidence level
if $|\xi|>0.074$.

\section{Conclusions}
\label{sec5}

We studied the cosmological consequences of Lorentz-violating
neutrinos.
We obtained the generalized Lagrangian for a neutrino with
the Coleman-Glashow type dispersion relation, and then
derived its Boltzmann equation and initial conditions in the
linear theory of cosmological perturbations.
Using the 7-year WMAP data in combination with measurements of
the Hubble constant and the BAO feature, we found no evidence
for Lorentz invariant violence in the neutrino sector.
The ongoing Planck experiment is expected to able to give
a more stringent constraints on the deformation parameter.

\acknowledgments Our numerical analysis was performed on the Lenovo
DeepComp 7000 supercomputer in SCCAS. This work is partially
supported by the two projects of Knowledge Innovation Program of
Chinese Academy of Science, NSFC under Grant No.11175225 and
No.10975167, and National Basic Research Program of China under
Grant No.2010CB832805 and No.2010CB833004. We used CosmoMC and CAMB.
We also acknowledge the use of WMAP data.

\end{document}